# The Faraday induction law in relativity theory

**A L Kholmetskii**
Department of Physics, Belarus State University,
4, F. Skorina Avenue, 220080 Minsk Belarus

**O V Missevitch**
Institute for Nuclear Problems, Belarus State University,
11, Bobruiskaya Str., 220050 Minsk, Belarus

**Abstract**.
We analyze the transformation properties of Faraday's law in an empty space and its relationship with Maxwell's equations. In our analysis we express the Faraday's law via the four-potential of electromagnetic field and the field of four-velocity, defined on a circuit under its deforming motion. The obtained equations show one more facet of the physical meaning of electromagnetic potentials, where the motional and transformer parts of the flux rule are incorporated into a common phenomenon, reflecting the dependence of four-potential on spatial and time coordinates, correspondingly. It has been explicitly shown that for filamentary closed deforming circuit the flux rule is Lorentz-invariant. At the same time, analyzing a transformation of e.m.f., we revealed a controversy: due to causal requirements, the e.m.f. should be the value of fixed sign, whereas the Lorentz invariance of flux rule admits the cases, where the e.m.f. might change its sign for different inertial observers. Possible resolutions of this controversy are discussed.

# 1. Introduction

It is known that Faraday's law involves two different phenomena in a single equation: law of induction ("transformer" part of induction), and Lorentz force law ("motional" part of induction, when the circuit moves) [1]. Numerous papers and books were devoted to Faraday's law. However, its transformation properties were analyzed up to date only in a few papers (see, *e.g.*, refs. [2-6]). The case of time-variable contour $\Gamma=\Gamma(t)$ happens to be especially difficult for such an analysis, and some ambiguities continue to exist up to now. In particular, one can mention the questions as follows:

1. Is the empirically discovered Faraday's law

$$\varepsilon = -\frac{d}{dt}\int_{S(t)} \vec{B}\cdot d\vec{S} \qquad (1)$$

equivalent to Maxwell's equation

$$\nabla \times \vec{E} = -\frac{\partial \vec{B}}{\partial t}, \qquad (2)$$

for the deforming contour $\Gamma(t)$, restricting the area $S(t)$? Here $\vec{E}$, $\vec{B}$ are the electric and magnetic fields, respectively, $\varepsilon$ is the e.m.f.

$$\varepsilon = \oint_{\Gamma(t)} \vec{f}\cdot d\vec{l}, \qquad (3)$$

and $\vec{f}$ is the force per unit charge.

2. If the answer to this question is positive, then the Faraday induction law (1) is necessarily Lorentz-invariant. In the case of negative answer, we have to analyze separately transformation properties of this empirical law.

3. Is there a common physical origin for "transformer" and "motional" parts of induction, joining them into a single law (1)? And what is a relationship between both parts of induction under space-time transformations?

4. Can we understand the known exceptions from the flux rule?

In order to find the answers to the listed questions, we suggest a new methodological approach, where Faraday's law is expressed through the four-potential of electromagnetic field. We believe that such an approach allows one to analyze the transformation properties of this law in more detail in comparison with the standard way, where the Faraday's law operates with the electric and magnetic fields, composing the components of the electromagnetic tensor.

# 2. On a relationship between Maxwell equation $\nabla \times \vec{E} = -\partial\vec{B}/\partial t$ and Faraday's law

It is known that the Faraday induction law was discovered experimentally in the first half of 19[th] century, and it suggested to Maxwell one of his equations, which was written by Hertz in its modern form (2). Now we ask a question: can we directly derive the Faraday's law (1) from Maxwell's equation (2)?

The answer to this question is trivial, when a closed line of integration is fixed in space. Then both $\Gamma$ and $S$ do not depend on time, and integrating Eq. (2) over $S$ and applying the Stokes theorem, we obtain

$$\int_S (\nabla\times\vec{E})d\vec{S} = -\int_S \frac{\partial\vec{B}}{\partial t}d\vec{S}, \quad \oint_\Gamma \vec{E}\cdot d\vec{l} = -\int_S \frac{\partial\vec{B}}{\partial t}\cdot d\vec{S}, \quad \varepsilon = -\frac{d}{dt}\int_S \vec{B}\cdot d\vec{S},$$

which is Eq. (1).

If a rigid contour Γ uniformly moves in space at the velocity $\vec{v}$, we can write [7]

$$\frac{d}{dt'}\int_S \vec{B}' \cdot d\vec{S}' = \int_S \frac{\partial \vec{B}'}{\partial t'} \cdot d\vec{S}' - \int_S \left[\nabla \times (\vec{v} \times \vec{B}')\right] d\vec{S}',$$

that yields (using $\frac{\partial \vec{B}}{\partial t} = -\nabla \times \vec{E}$ and applying Stokes theorem to *rhs*):

$$\oint_\Gamma (\vec{E}' + \vec{v} \times \vec{B}') d\vec{l}' = -\frac{d}{dt'}\int_S \vec{B}' \cdot d\vec{S}'. \tag{4}$$

Considering the latter expression, the authors of ref. [2] suggested interpreting the integrand of *lhs* as the electric field *E* in the rest frame of circuit K. Then the equality

$$\vec{E} = \vec{E}' + \vec{v} \times \vec{B}' \tag{5}$$

obviously disagrees with the relativistic transformation of an electric field. However, such an interpretation is erroneous, because both integrals in Eq. (4) should be evaluated at different time moments for different points of circuit, according to the rule for computing retarded e.m.f. (see below). Therefore, Eq. (5) is incorrect, since the field $\vec{E}'$ and $\vec{B}'$ should be determined for the same fixed instant *t* of the rest frame of the circuit, which corresponds to different *t'*.

The case, when a circuit moves and simultaneously changes its shape, is even more complicated, and the authors of ref. [2] concluded that the flux rule loses its simple relation with the Maxwell equation (2).

Further contributions [3, 4] emphasized incorrectness of Eq. (5) in case of relativistic velocities $\vec{v}$. For such a case Monsivais suggested to modify the Faraday's law into the form [4]

$$-\frac{1}{c}\frac{d}{dt}\iint_{S(t)} B(\vec{r},t) \cdot \vec{n} d\sigma = \oint_{\Gamma'} \frac{\vec{E}'(\vec{r}',t')}{\gamma'(\vec{r}',t')} \cdot d\vec{l}', \tag{6}$$

where

$\vec{B}, \vec{E}$ are the magnetic and electric fields, respectively,

$d\sigma$ is the element of the area of the surface *S*, restricted by the contour Γ,

$d\vec{l}$ is the element of the contour Γ,

$\gamma' = \sqrt{1 - v^2/c^2}$,

*c* is the light velocity in vacuum,

and primed variables refer to the Lorentz frame K', co-moving with an arbitrary point of the circuit at an arbitrary instant, and $\vec{v}$ is the relative velocity between K and K'. The spatial $\vec{r}'$ and time *t'* coordinates refer to the frames K', and the values of *t'* are such that *t* is a fixed value along the trajectory. Following the notation of ref. [4], we have used a dotted circle on the right integral, which denotes the fact that each its infinitesimal term is measured in different frames K'.

This equation was first obtained by Gelman [3]. The author of ref. [4] invented a special symbol: dotted circle in Eq. (6) to indicate that an integral over Γ belongs to a special kind. Eq. (6) can be considered as the consequence of Maxwell equation (2) and relativistic transformation of electromagnetic field. However, the authors of ref. [3, 4] concluded that a direct calculation of e.m.f. in Eq. (6) is not a simple problem in a general case. Even for rigidly moving circuit, when a well-defined e.m.f. *ε'* in K' exists,

$$\varepsilon' = \oint_{\Gamma'} \vec{E}'(\vec{r}',t') \cdot d\vec{l}', \tag{7}$$

we cannot simply put $\varepsilon(t) = \varepsilon'(t')/\gamma$, because a calculation of e.m.f.

$$\varepsilon = \frac{1}{\gamma'} \oint_{\Gamma'} \vec{E}'(\vec{r}',t') \cdot d\vec{l}' \tag{8}$$

is carried out at a fixed $t$ and the one in Eq. (7) at a fixed $t'$ [4]. Monsivias mentions [4] that for the curve $\Gamma(t)$ as a finite union of curves

$$\Gamma(t) = \bigcup_{i=1}^{n} \Gamma_i(t), \tag{9}$$

where each $\Gamma_i(t)$ moves with constant velocity $\vec{v}_i$, the integral of the *rhs* of Eq. (6) becomes a sum of standard integrals, and e.m.f. can be calculated analytically for the particular forms of Eq. (9).

Thus, for a deforming contour, the Faraday's law occurs very complicated, when it is expressed via the electric and magnetic fields, and it loses a simple relationship with the Maxwell equation (2). The author of Ref. [5] even suggests modifying Maxwell's equations to make them compatible with the flux rule in a moving frame. In our opinion, an actual further progress in the analysis of Faraday's law can be achieved, when we express the Faraday law through the vector and scalar potentials of electromagnetic field.

### *2.1. The Faraday induction law expressed via the potentials of electromagnetic field*

As a starting point, we write the Faraday induction law in a well-known explicit form

$$\oint_{\Gamma(t)} \left( \vec{E}(\vec{r},t) + \vec{u}(\vec{r},t) \times \vec{B}(\vec{r},t) \right) d\vec{l} = -\frac{d}{dt} \int_{S(t)} \vec{B}(\vec{r},t) \cdot d\vec{S}, \tag{10}$$

with clear indication of the dependence of $\vec{E}, \vec{B}$ and $\vec{u}$ on space-time coordinates. Here $\vec{u}(\vec{r},t)$ is the velocity of point $\vec{r} \in \Gamma(t)$ at the instant $t$. (In general, the velocity vector field $\vec{u}(\vec{r},t)$ can be introduced throughout the area $S(t)$, too [4]). The *lhs* of this equation represents an e.m.f., according to its general definition (3) and the Lorentz force law. The *rhs* of Eq. (10) describes a time rate of the magnetic flux across the surface $S$ capping the circuit $\Gamma$. Here we assume that Eq. (10) is written in the rest frame K of a measuring instrument (voltmeter), and both integrals are computed at the same time moment. We also assume that all functions in Eq. (10) are continuous.

Now let us express the Faraday law (10) via the scalar $\varphi$ and vector $\vec{A}$ potentials of the electromagnetic field, using their relationships with $\vec{E}$ and $\vec{B}$:

$$\vec{E} = -\frac{\partial \vec{A}}{\partial t} - \nabla \varphi, \ \vec{B} = \nabla \times \vec{A}. \tag{11}$$

Combining Eqs. (10) and (11), we get the following relationship:

$$\oint_{\Gamma(t)} \left( -\frac{\partial \vec{A}}{\partial t} - \nabla \varphi + \vec{u} \times (\nabla \times \vec{A}) \right) d\vec{l} = -\frac{d}{dt} \int_S (\nabla \times \vec{A}) d\vec{S}. \tag{12}$$

For further transformation of Eq. (12) we use the vector identities

$$\frac{d\vec{A}}{dt} = \frac{\partial \vec{A}}{\partial t} + (\vec{u} \cdot \nabla)\vec{A}, \tag{13}$$

$$\vec{u} \times (\nabla \times \vec{A}) = \nabla_A (\vec{u} \cdot \vec{A}) - (\vec{u} \cdot \nabla)\vec{A}, \tag{14}$$

where we should be careful in application of operator $\nabla$. Namely, we take into account that the velocity vector field $\vec{u}$ is a function of $\vec{r} \in \Gamma$, and following Feynman [1], introduce a partial operator $\nabla_A$ in the latter identity (14), where the subscript denotes the quantity to be differentiated. We also notice that the circular integral of a total gradient is equal to zero:

$$\oint_{\Gamma(t)} \nabla\varphi \cdot d\vec{l} = 0, \quad (15)$$

as well as apply the Stokes theorem

$$\int_S (\nabla \times \vec{A})d\vec{S} = \oint_\Gamma \vec{A} \cdot d\vec{l} \quad (16)$$

to *rhs* of Eq. (12). Combining Eqs. (12)-(16), we arrive at

$$\oint_{\Gamma(t)} \left(-\frac{d\vec{A}}{dt} + \nabla_A(\vec{u}\cdot\vec{A})\right)d\vec{l} = -\frac{d}{dt}\oint_{\Gamma(t)} \vec{A}\cdot d\vec{l}. \quad (17)$$

For further transformation of *rhs* of Eq. (17), we have to evaluate the total time derivative of the integral $\oint_{\Gamma(t)} \vec{A}\cdot d\vec{l}$. One can show that

$$\frac{d}{dt}\oint_{\Gamma(t)} \vec{A}\cdot d\vec{l} = \oint_{\Gamma(t)} \left(\frac{d\vec{A}}{dt} + \vec{A}\times(\nabla\times\vec{u}) + (\vec{A}\cdot\nabla)\vec{u}\right)d\vec{l},$$

or in a compact form

$$\frac{d}{dt}\oint_{\Gamma(t)} \vec{A}\cdot d\vec{l} = \oint_{\Gamma(t)} \left(\frac{d\vec{A}}{dt} + \nabla_u(\vec{u}\cdot\vec{A})\right)d\vec{l}.$$

A proof of this theorem is straightforward (see Appendix A).

Using this theorem, we transform Eq. (17) into

$$\oint_{\Gamma(t)} \left(-\frac{d\vec{A}}{dt} + \nabla_A(\vec{u}\cdot\vec{A})\right)d\vec{l} = \oint_{\Gamma(t)} \left[-\frac{d\vec{A}}{dt} - \nabla_u(\vec{u}\cdot\vec{A})\right]d\vec{l}. \quad (18)$$

We see that Eq. (18) represents a mathematical identity, because $\oint_\Gamma \nabla_A(\vec{u}\cdot\vec{A})d\vec{l} + \oint_\Gamma \nabla_u(\vec{u}\cdot\vec{A})d\vec{l} = \oint_\Gamma \nabla(\vec{u}\cdot\vec{A})d\vec{l} = 0$, as a circular integral of the total gradient, and

$$\oint_\Gamma \nabla_A(\vec{u}\cdot\vec{A})d\vec{l} = -\oint_\Gamma \nabla_u(\vec{u}\cdot\vec{A})d\vec{l}. \quad (19)$$

Now we are able to establish a relationship between Maxwell's equations and Faraday's law in full. Namely, the latter law cannot be obtained from Eq. (2) exclusively: its derivation for an arbitrary deforming contour requires involving the Lorentz force law and the solutions (11) of a couple of Maxwell's equation ($\nabla\times\vec{E} = -\frac{\partial\vec{B}}{\partial t}$, $\nabla\cdot\vec{B} = 0$) giving the relationship between electric and magnetic fields. Therefore, the flux rule should be inevitably Lorentz-invariant. At the same time, it seems important to show explicitly the invariance of Faraday' law (18) for better understanding of its physics.

## 3. Transformation properties of the Faraday induction law

It is worth to mention at the beginning of this section that the Lorentz invariance of Faraday's law (flux rule) has been already shown in a general form [6], where the author suggested a new Lorentz-invariant generalized form for a magnetic flux. At the same time, such a general covariant formulation of Faraday's law makes difficult to apply it to analysis of many particular physical problems, dealing with computing an e.m.f. and magnetic fluxes. In addition, the authors of the cited above papers did not stress an essential feature of Faraday's law transformation in comparison with many other relativistic problems. It is related to physical interpre-tation of the Lorentz transformations, suggested by Einstein in his fundamental paper [8]. Namely, if two inertial frames K and K' are in relative motion, that each observer in his own rest frame uses his own measuring instrument to determine physical quantities in another frame. However, one can see that this is often not the case for the Faraday induction law: a measuring instrument for e.m.f. (voltmeter) usually represents an inherent part of the moving circuit, and hence, any inertial observer, regardless of his particular velocity with respect to the circuit, uses this voltmeter in his measurements. In principle, one can demand that a moving observer operates with his own voltmeter, included into a circuit by means of sliding contacts. Obviously, the problem, where all inertial observers use in their measurements a single voltmeter, integrated into a circuit, differs from the problem, where each observer uses his own voltmeter. Since the latter case has no practical significance, we will analyze the transformation properties of the Faraday induction law, where all observers use the same measuring instrument. Then only in the rest frame of voltmeter K a circular integration over $\Gamma$ is carried out at the same instant $t$, determining $\varepsilon(t)$. There is no physical meaning to integrate over the circuit at a fixed moment $t'$ of some arbitrary inertial frame K' to find $\varepsilon'(t')$, because this value has no simple relation with an actual indication of voltmeter. In order to find such a relationship, an integration time $t'$ should be connected with $t$ by means of the Lorentz transformation

$$t' = \gamma(t - \vec{v}\vec{r}/c^2). \tag{20}$$

for $t = \mathrm{const}$. Here $\vec{v}$ stands for the velocity of voltmeter in the frame K', and $\vec{r}$ belongs to the closed circuit in the rest frame of voltmeter K. One sees from Eq. (20) that the time moments $t'$ are different for different $\vec{r}$. This rule, where $t' = t'(\vec{r}, \vec{v})$, was named by Cullwick [9] as a rule for computing retarded (advanced) e.m.f. It appears due to the above-mentioned fact: a measuring instrument (voltmeter) is common for all inertial observers. Of course, such a voltmeter can be placed at an arbitrary point of a deforming circuit. However, as soon as we choose this point and its rest frame K, further integration over a circuit in any Lorentz frame K' should be carried out under the requirement $t$=const in the frame K.

In order to explore the transformation properties of Faraday's law, we have to generalize Eq. (18) into a four-dimensional form, introducing the four-potential $A^\alpha(\varphi, \vec{A})$ and the field of four-velocity $u^\alpha \{\gamma_u, \gamma_u \vec{u}/c\}$ ($\alpha$=0...3, $\gamma_u = 1/\sqrt{1 - u^2/c^2}$). For this purpose we use the equalities

$$\frac{dA_i}{dt} = c \frac{\partial A_i}{\partial x^\alpha} \frac{u^\alpha}{\gamma_u} \quad (x^0 = ct, i = 1..3), \tag{21}$$

(see, Eq. (13)),

$$\left[ c\nabla \cdot \varphi - \nabla_A (\vec{u} \cdot \vec{A}) \right]_i = c \frac{u^\alpha}{\gamma_u} \frac{\partial A_\alpha}{\partial x^i}, \text{ and} \tag{22}$$

$$\left[ \nabla_u (\vec{u} \cdot \vec{A}) \right]_i = c A_\alpha \frac{\partial}{\partial x^i} \frac{u^\alpha}{\gamma_u}. \tag{23}$$

We also take into account Eq. (15), which is applied to the rest frame of the voltmeter. Then, combining Eqs. (18), (21)-(23), we arrive at

$$\oint_\Gamma \left( \frac{\partial A_i}{\partial x^\alpha} \frac{u^\alpha}{\gamma_u} - \frac{u^\alpha}{\gamma_u} \frac{\partial A_\alpha}{\partial x^i} \right) dx^i = \oint_\Gamma \left( \frac{\partial A_i}{\partial x^\alpha} \frac{u^\alpha}{\gamma_u} + A_\alpha \frac{\partial}{\partial x^i} \frac{u^\alpha}{\gamma_u} \right) dx^i. \tag{24}$$

Mathematical correctness of Eq. (24) is proved by the equality

$$\oint_\Gamma \left( \frac{u^\alpha}{\gamma_u} \frac{\partial A_\alpha}{\partial x^i} + A_\alpha \frac{\partial}{\partial x^i} \frac{u^\alpha}{\gamma_u} \right) dx^i = \oint_\Gamma \frac{\partial}{\partial x^i} \frac{A_\alpha u^\alpha}{\gamma_u} dx^i = 0, \qquad (25)$$

as far as the integral is computed for fixed $t$ in the rest frame of the voltmeter. Then

$$\oint_\Gamma \frac{u^\alpha}{\gamma_u} \frac{\partial A_\alpha}{\partial x^i} dx^i = -\oint_\Gamma A_\alpha \frac{\partial}{\partial x^i} \frac{u^\alpha}{\gamma_u} dx^i, \qquad (26)$$

which, in turn, provides the equality (24).

As in Eq. (18), the *lhs* of Eq. (24) describes an e.m.f. Using a tensor of the electromagnetic field $F_{\alpha\beta} = \frac{\partial A_\beta}{\partial x^\alpha} - \frac{\partial A_\alpha}{\partial x^\beta}$, we can present the e.m.f. in form

$$\varepsilon = \oint_\Gamma \left( \frac{\partial A_i}{\partial x^\alpha} \frac{u^\alpha}{\gamma_u} - \frac{u^\alpha}{\gamma_u} \frac{\partial A_\alpha}{\partial x^i} \right) dx^i = -\oint_\Gamma \frac{F_{i\alpha} u^\alpha}{\gamma_u} dx^i, \qquad (27)$$

where $F_{\alpha\beta} u^\beta$ is the four-dimensional Lorentz force per unit charge [10]. That is why we may consider Eq. (27) as a generalization of definition of e.m.f. (3).

Further, continuing consideration of the Faraday's law transformation, we have, in general, to distinguish two separate problems: implementation of the flux rule for different observers, and transformation of e.m.f. itself. These problems are consequently analyzed in sub-sections 3.1 and 3.2

### 3.1. Flux rule for different observers

Eq. (24) can be used to explore transformation properties of the Faraday induction law (flux rule). Note that under derivation of Eq. (24) we did not make any special assumptions about the behavior of the closed circuit $\Gamma$ in the rest frame of the voltmeter K. Hence, both integrands in *lhs* and *rhs* of Eq. (24) have the same form for primed functions, defined in an arbitrary inertial frame K'. Then the Lorentz invariance of flux-rule would mean the equality

$$\oint_{\Gamma'} \left( \frac{\partial A'_i}{\partial x'^\alpha} \frac{u'^\alpha}{\gamma'_u} - \frac{u'^\alpha}{\gamma'_u} \frac{\partial A'_\alpha}{\partial x'^i} \right) dx'^i = \oint_{\Gamma'} \left( \frac{\partial A'_i}{\partial x'^\alpha} \frac{u'^\alpha}{\gamma'_u} + A'_\alpha \frac{\partial}{\partial x'^i} \frac{u'^\alpha}{\gamma'_u} \right) dx'^i, \qquad (28)$$

which is possible, if

$$\oint_{\Gamma'} \frac{\partial}{\partial x'^i} \frac{A'_\alpha u'^\alpha}{\gamma'_u} dx'^i = 0 \qquad (29)$$

(compare with Eq. (25)). In Eqs. (28) and (29) the rule for computing of a retarded e.m.f. should be applied. This means that the four-vectors depend not on $x'^\alpha$, but on $(x'^i, x'^0(x^i))$ in accordance with Eq. (20). Hence we can designate

$$\oint_{\Gamma'} \frac{\partial}{\partial x'^i} \frac{A'_\alpha u'^\alpha}{\gamma'_u} dx'^i = \oint_{\Gamma'} \frac{\partial}{\partial x'^i} G(x'^i, x'^0(x^i)) dx'^i,$$

where $G = \frac{A'_\alpha u'^\alpha}{\gamma'_u}$. Then we obtain

$$\oint_{\Gamma'} \frac{\partial}{\partial x'^i} G(x'^i, x'^0(x^i)) dx'^i = \oint_{\Gamma'} \left( \frac{\partial G}{\partial x'^i} dx'^i + \frac{\partial G}{\partial x'^0} \frac{\partial x'^0}{\partial x^j} \frac{\partial x^j}{\partial x'^i} dx'^i \right) = \oint_{\Gamma'} dG = 0,$$

as a circular integral of a full differential. This proves the validity of Eq. (29), and hence, the validity of Eq. (28).

Thus, the flux rule is implemented for any inertial observer. Our conclusion coincides with the results of refs. [3, 4, 6], although our proof of the Lorentz-invariance of flux rule ex-pressed via four-potential is the simplest among those presented to date and does not require a knowledge of the explicit form of magnetic flux transformation law, or introducing a new physical quantity such as 4-flux [6].

A notation of Faraday's law in forms (18) and (24) seems to be novel. Perhaps, Eq. (24) is not significant for practical applications of Faraday's law, although it clearly indicates, in combination with the rule for computing a retarded (advanced) e.m.f., the Lorentz-invariance of this law.

The obtained equations allow revealing a physical meaning of the transformer and motional parts of Faraday's law in terms of four-potential of electromagnetic field: namely, the transformer part is determined by the change of $\vec{A}$ with $t$, while the motional part is determined by the change of $\vec{A}$ with $\vec{r}$. Thus, both parts of Faraday's law have a common physical origin: variation of the vector potential as the function of space-time four-vector. Of course, under space-time transformations a partial time derivative of vector field $\vec{A}$ can be transformed into its partial spatial derivative, and vise versa. Hence, a relative contribution of the motional and transformer parts of Faraday's law into the resultant e.m.f. is frame-dependent, and in the limiting particular cases the e.m.f. in one frame can be purely transformer, while in another frame it can be purely motional. Einstein specially emphasized this circumstance in his famous work [8]. Now we can say that both parts of Faraday's law have a common physical origin: variation of $\vec{A}$ as the function of space-time four-vector. A visible distinction between two parts of Faraday's law appears at the level, when we express this law through the electric and magnetic fields. Even so, we have to remember about a common origin of these phenomena.

We would like to point out that our analysis was carried out for the geometrical closed lines or filamentary closed circuits. In practice we can deal with the closed circuits, which include extended conductors (e.g., an unipolar inductor). In contrast to filamentary closed lines, such a circuit can contain the points of discontinuity of a velocity vector field (*e.g.*, in the sliding contacts on the boundary of moving extended conductors). Then the identities of vector analysis, containing a velocity parameter (for example, Eqs. (13), (14)) become, in general, incorrect, and the flux rule can be violated, too. At the same time, for such distributed circuits one can choose a filamentary closed line, where the points of discontinuity of a velocity vector field are avoided, and the flux rule is fulfilled (see [11] and references therein). Then it is Lorentz invariant according to our analysis.

*3.2. Transformation of an e.m.f.*

Analyzing the transformation of an e.m.f. itself, we are free to take an e.m.f. for the rest frame of the voltmeter in any available form (see, *lhs* of Eqs. (10), (18) and Eq. (25)) and then compute its value in any other Lorentz frame. For this purpose we have to transform the velocity vector field $\vec{u}$, and then to apply either a transformation of electromagnetic fields, or electromagnetic potentials, in accordance with the chosen form of Faraday's law. In addition, we have to determine an integration time in accordance with Eq. (9). It is important that in such calculations we deal with only two Lorentz frames: the rest frame of voltmeter K and external observer K', although a special integration procedure of Eq. (5) can be also applied. Nevertheless, one should recognize that for an arbitrarily moving deformed circuit any available approach (including a parametrical integration of Ref. [6]) does not allow analytic derivation of the law of transformation of e.m.f. In these conditions the causal requirements give a key to predicting a form of e.m.f. transformation. Indeed:
- if an e.m.f. is equal to zero in one frame of observation, that it should be equal to zero in any other frame,

- if an e.m.f. is positive (negative) in one inertial frame, that it should be positive (negative) in any other frame,
- for $v=0$ (observer be at rest with respect to voltmeter), $\varepsilon = \varepsilon'$.

One can see that these three requirements, taken together, determine a form of transformation of e.m.f. at the lowest order in $1/c$ as follows:

$$\varepsilon' = \varepsilon \cdot F(v^2/c^2), \text{ with } F(0) = 1, \qquad (30)$$

where $\varepsilon$ is the e.m.f. in a rest frame of voltmeter K, $\varepsilon'$ is the e.m.f. in an external Lorentz frame K', and $\vec{v}$ is the relative velocity of K and K'.

Cullwick considered a number of simple problems, where he derived $F(v^2/c^2) = \gamma$ [9]. The same form of function $F$ can be obtained for any moving deforming circuit, if we assume the simplest construction of a voltmeter, common for all inertial observers: a resistor with a high resistance $R$, connected in series with galvanometer, measuring a current through $R$. Then the e.m.f. is equal to

$\varepsilon = RI$.

For any inertial frames K and K', moving at a relative velocity $\vec{v}$, $R'=R$, and $I=I'/\gamma$ (see, Ref. [9]). Hence,

$$\varepsilon' = I'R' = \gamma IR = \gamma \varepsilon. \qquad (31)$$

At the same time, there is a serious contradiction between Eqs. (30), (31) and the proven above the Lorentz invariance of flux rule. It appears in the problems, where the magnetic field $\vec{B}$ is constant or very slowly changes with time and distance, so that its spatial and time partial derivatives are negligible. Then the Faraday's law acquires the form (see. Eq. (1))

$$\varepsilon = -\frac{d}{dt} \int_{S(t)} \vec{B} \cdot d\vec{S} = -\vec{B} \cdot \frac{d\vec{S}}{dt}.$$

Certainly, the time derivative $d\vec{S}/dt$ is the value of fixed sign in all inertial reference frames. However, the magnetic field $\vec{B}$, constituting the components of the tensor of EM field, can change its sign for different observers. In particular, it happens when the field invariants satisfy the requirements

$$\vec{E} \cdot \vec{B} = 0, \ E^2 - c^2 B^2 > 0. \qquad (32)$$

Then the whole product $\vec{B} \cdot d\vec{S}/dt$ is an alternating quantity for different inertial observers. In other words, if in an inertial frame K

$$\varepsilon = -\vec{B} \cdot \frac{d\vec{S}}{dt} > 0, \qquad (33)$$

then we can find another inertial frame K', where

$$\varepsilon' = -\vec{B}' \cdot \frac{d\vec{S}'}{dt'} < 0, \qquad (34)$$

due to the Lorentz invariance of flux rule and different signs of $\vec{B}$ and $\vec{B}'$. Hence according to Eqs. (32), (33) an e.m.f. represents an alternating quantity. Thus, Eq. (31) and Eqs. (32), (33), all obtained as unambiguous consequences of classical electrodynamics, contradict each other.

Analyzing the revealed controversy, we assume the constant magnetic field within a closed circuit $\Gamma$ (the case of Eqs. (33), (34)). Then the requirement (32) signifies the existence of such a Lorenz frame, wherein the magnetic field $\vec{B}$ is identically equal to zero in a vicinity of $\Gamma$, and only the electric field $\vec{E}$ exists. This field does not penetrate into a bulk of conducting circuit, where the internal electric field $\vec{E}_{in} = 0$. Physically it means that a conductor is po-larized in such a way, where its internal electric field is always vanished. Insofar the magnetic field is equal to zero in this frame of observation, we obtain the equality

$$\vec{E}_{in}, \vec{B}_{in} = 0. \qquad (35)$$

everywhere inside the conductor. Due to linearity of the electromagnetic field transformation, we get

$$\vec{E}'_{in}, \vec{B}'_{in} = 0 \qquad (36)$$

in any other Lorentz frame of observation. Hence the e.m.f. is equal to zero in all frames of observation in accordance with Eqs. (30), (31). In fact, the revealed incompatibility of Eq. (30), (31) and Eqs. (33), (34) signifies a presence of one more exception from the flux rule, which was not reported up to date. It can be demonstrated by the problem in Fig. 1.

An expanding conducting circumference with the radius $R(t)=R(0)+vt$ is placed inside the parallel plate condenser. In the rest frame of condenser, only the constant electric field $E$ exists along the axis $y$, and we get Eq. (35) for the internal electric and magnetic field of the conductor. Now let us introduce an inertial frame K, wherein the condenser and circumference move at the constant velocity $u$ along the axis $x$. Since in this frame Eq. (36) holds true, then there is no e.m.f. in the circuit. At the same time, the magnetic flux $\Phi$ across the area of expanded circumference $S$ and its total time derivative $d\Phi/dt$ are not vanished in the frame K. Indeed, in this frame

$$B_z = \frac{uE}{c^2 \sqrt{1 - u^2/c^2}},$$

between the plates of condenser, and

$$\frac{d\Phi}{dt} = \frac{dS}{dt} B_z = 2\pi R(t) \frac{dR(t)}{dt} B_z \approx 2\pi R(t) \frac{uvE}{c^2} \qquad (37)$$

to the accuracy $c^{-2}$. Thus, we observe that the time derivative of the magnetic flux across the circuit in Fig. 1 is not vanished, while the e.m.f. in the circuit is equal to zero. We emphasize that the revealed exception from the flux rule happens for a filamentary circuit, when the velocity field is a continuous function. At the same time, the electric and magnetic fields contain the points of discontinuity on a charged surface of conductor, and the flux rule is no longer valid. These points of discontinuity define a principal difference between real conducting filamentary circuits and closed geometrical lines in space. For geometrical lines the controversy between Eqs. (30), (31) and Eqs. (33), (34) persists, and this puzzling requires to be carefully analyzed.

## 4. Conclusions

In the introduction to this paper we formulated a number of questions, and all of them are answered below:
1. Is the empirically discovered Faraday's law (1) equivalent to Maxwell's equation (2) for the deforming contour $\Gamma(t)$, restricting the area $S(t)$?

In general, it is not. **In order to derive Faraday's law we have to involve the Lorentz force law and a couple of Maxwell's equations** $\nabla \times \vec{E} = -\frac{\partial \vec{B}}{\partial t}$, $\nabla \cdot \vec{B} = 0$.

2. Is the flux rule Lorentz-invariant?

**We explicitly showed the invariance of flux rule, expressed through a four-potential of electromagnetic field.**

3. Is there a common physical origin for "transformer" and "motional" parts of induction, joining them into a single law (1). And what is a relationship between both parts of induction under space-time transformations?

**Expressing the Faraday law through the four-potential, we found that the transformer part of Faraday's is determined by the change of $\vec{A}$ with *t*, while the motional part is determined by the change of $\vec{A}$ with $\vec{r}$. Therefore, both parts of Faraday's law can be incorporated into a common physical phenomenon: variation of $\vec{A}$ as the function of space-time four-vector.** Under space-time transformations a partial time derivative of vector field $\vec{A}$ can be transformed into its partial spatial derivative, and vise versa. Hence, a relative contribution of the motional and transformer parts of Faraday's law into the resultant e.m.f. is the frame-dependent quantity.

4. Can we understand the known exceptions from the flux rule?

The known exceptions from the flux rule are related with the properties of the closed circuits (filamentary lines or extended conductors), and do not depend on the properties of electromagnetic fields and four-potentials. Therefore, both forms of Faraday law (expressions through the fields and through potentials) are relevant for analysis of these exceptions.

At the same time, finding the answers on the foregoing questions, we faced the challenge, representing new serious difficulty of classical electrodynamics: **the e.m.f. transformations (30) and (31), being in conformity with causal requirements, contradict the Lorentz invariance of flux rule, where an e.m.f. might be an alternating quantity (Eqs. (33), (34)).** We have shown that for real conducting circuits this controversy is reduced to one more exception from the flux rule. In contrast to the known up to date exceptions, happened for extended conductors, the new exception takes place for filamentary conducting closed circuits, and it removes the contradiction between Eqs. (30), (31) and Eqs. (33), (34). However, for closed geometrical lines in space the controversy between Eqs. (30), (31) and Eqs. (33), (34) persists, and this puzzling requires to be carefully analyzed.

**Appendix A**

In this appendix we prove a theorem as follows:

- for any smooth vector field $\vec{A}(\vec{r},t)$ and any smooth closed line $\Gamma(t)$,

$$\frac{d}{dt}\oint_{\Gamma(t)}\vec{A}\cdot d\vec{l} = \oint_{\Gamma(t)}\left(\frac{d\vec{A}}{dt}+\vec{A}\times(\nabla\times\vec{u})+(\vec{A}\cdot\nabla)\vec{u}\right)d\vec{l}, \tag{A1}$$

where $\vec{u}=\vec{u}(\vec{r},t)$ is the velocity of the point $\vec{r}$ within the line $\Gamma$.

**Proof:** Let us write the integral $\oint_{\Gamma(t)}\vec{A}\cdot d\vec{l}$, proceeding from its definition:

$$\oint_{\Gamma(t)}\vec{A}\cdot d\vec{l} = \oint_{\Gamma(t)}(A_x(\vec{r},t)dx + A_y(\vec{r},t)dy + A_z(\vec{r},t)dz) = \lim_{n\to\infty}\sum_{i=1}^{n}(A_x(\vec{r}_i,t)\Delta x_i + A_y(\vec{r}_i,t)\Delta y_i + A_z(\vec{r}_i,t)\Delta z_i),$$

where $\Delta x_i = x_i - x_{i-1}$, $\Delta y_i = y_i - y_{i-1}$, $\Delta z_i = z_i - z_{i-1}$ are the components of elemental lines on the corresponding coordinate axes, and $\vec{r}_i$ is the coordinate of any point on the elemental line *i*. Then

$$\frac{d}{dt}\oint_{\Gamma(t)}\vec{A}\cdot d\vec{l} = \lim_{n\to\infty}\left[\frac{d}{dt}\sum_{i=1}^{n}(A_x(\vec{r}_i,t)\Delta x_i + A_y(\vec{r}_i,t)\Delta y_i + A_z(\vec{r}_i,t)\Delta z_i)\right] =$$

{designating $A_x(\vec{r}_i,t)=\vec{A}_{xi}$, $A_y(\vec{r}_i,t)=\vec{A}_{yi}$, $A_z(\vec{r}_i,t)=\vec{A}_{zi}$}

12$$= \lim_{n\to\infty} \sum_{i=1}^{n} \left( \frac{dA_{xi}}{dt}\Delta x_i + \frac{dA_{yi}}{dt}\Delta y_i + \frac{dA_{zi}}{dt}\Delta z_i + A_{xi}\left(\frac{dx_i}{dt} - \frac{dx_{i-1}}{dt}\right) + A_{yi}\left(\frac{dy_i}{dt} - \frac{dy_{i-1}}{dt}\right) + A_{zi}\left(\frac{dz_i}{dt} - \frac{dz_{i-1}}{dt}\right) \right) =$$

$$= \lim_{n\to\infty} \sum_{i=1}^{n} \left( \frac{dA_{xi}}{dt}\Delta x_i + \frac{dA_{yi}}{dt}\Delta y_i + \frac{dA_{zi}}{dt}\Delta z_i + A_x(u_{xi} - u_{xi-1}) + A_{yi}(u_{yi} - u_{yi-1}) + A_{zi}(u_{zi} - u_{yz-1}) \right) =$$

$$= \oint_{\Gamma(t)} \frac{d\vec{A}}{dt}(\vec{r},t) \cdot d\vec{l} + \lim_{n\to\infty} \sum_{i=1}^{n} \left( A_{xi}\Delta u_{xi} + A_{yi}\Delta u_{yi} + A_{zi}\Delta u_{zi} \right).$$

Further, taking into account that for any fixed time moment the velocities $\vec{v}_i$ are considered as the smooth functions of $\vec{r}_i$ within the closed line $\Gamma(t)$, we can write

$$\Delta u_{xi} = \frac{\partial u_x}{\partial x}\Delta x_i + \frac{\partial u_x}{\partial y}\Delta y_i + \frac{\partial u_x}{\partial z}\Delta z_i, \quad \Delta u_{yi} = \frac{\partial u_y}{\partial x}\Delta x_i + \frac{\partial u_y}{\partial y}\Delta y_i + \frac{\partial u_y}{\partial z}\Delta z_i,$$

$$\Delta u_{zi} = \frac{\partial u_z}{\partial x}\Delta x_i + \frac{\partial u_z}{\partial y}\Delta y_i + \frac{\partial u_z}{\partial z}\Delta z_i.$$

From there we obtain

$$\frac{d}{dt}\oint_{\Gamma(t)} \vec{A}\cdot d\vec{l} = \oint_{\Gamma(t)}\frac{d\vec{A}}{dt}(\vec{r},t)\cdot d\vec{l} + \oint_{\Gamma(t)} A_x\left(\frac{\partial u_x}{\partial x}dx + \frac{\partial u_x}{\partial y}dy + \frac{\partial u_x}{\partial z}dz\right) + \oint_{\Gamma(t)} A_y\left(\frac{\partial u_y}{\partial x}dx + \frac{\partial u_y}{\partial y}dy + \frac{\partial u_y}{\partial z}dz\right) +$$

$$+ \oint_{\Gamma(t)} A_z\left(\frac{\partial u_z}{\partial x}dx + \frac{\partial u_z}{\partial y}dy + \frac{\partial u_z}{\partial z}dz\right) =$$

$$= \oint_{\Gamma(t)}\frac{d\vec{A}}{dt}(\vec{r},t)\cdot d\vec{l} + \oint_{\Gamma(t)}\left(A_x\frac{\partial u_x}{\partial x} + A_y\frac{\partial u_y}{\partial x} + A_z\frac{\partial u_z}{\partial x}\right)dx + \left(A_x\frac{\partial u_x}{\partial y} + A_y\frac{\partial u_y}{\partial y} + A_z\frac{\partial u_z}{\partial y}\right)dy + \quad \text{(A2)}$$

$$+ \oint_{\Gamma(t)}\left(A_x\frac{\partial u_x}{\partial z} + A_y\frac{\partial u_y}{\partial z} + A_z\frac{\partial u_z}{\partial z}\right)dz = \oint_{\Gamma(t)}\frac{d\vec{A}}{dt}(\vec{r},t)\cdot d\vec{l} + \oint_{\Gamma(t)}\nabla_u(\vec{u}\cdot\vec{A})d\vec{l}$$

where the operator $\nabla_u$ acts only on $\vec{u}$. It can be expressed via the conventional operator $\nabla$ as

$$\nabla_u(\vec{u}\cdot\vec{A}) = \vec{A}\times(\nabla\times\vec{u}) + (\vec{A}\cdot\nabla)\vec{u}. \quad (A3)$$

Substituting its value into Eq. (A2) one gets Eq. (A1), which proves the theorem.

## References

1. Feynman R P 1964 *The Feynman Lectures on Physics* vol II (Reading, Mass: Addison-Wesley)
2. Ares-de-Parga G and Rosales M A 1989 Eur. J. Phys. **10** 74
3. Gelman H 1991 Eur. J. Phys. **12** 230
4. Monsivais G 2004 Am. J. Phys. **72** 1178
5. Dunning-Davies J  arxiv.org/abs/physics/0406056
6. Marx E 1975 J. Franklin Inst. **300** 353
7. Panofsky W K H and Phillips M 1962 *Classical Electricity and Magnetism* 2nd edn (Reading, Mass.: Addison-Wesley)
8. Einstein A 1905 Ann. Phys. **17** 891
9. Cullwick E A 1957 *Electromagnetism and Relativity* (London-New York-Toronto: Longmans, Green and Co)
10. Landau L D and Lifshitz E M 1962 *The Classical Theory of Fields*, 2nd edn (New York: Pergamon Press)
11. Munley F 2004 Am. J. Phys. **72** 1478

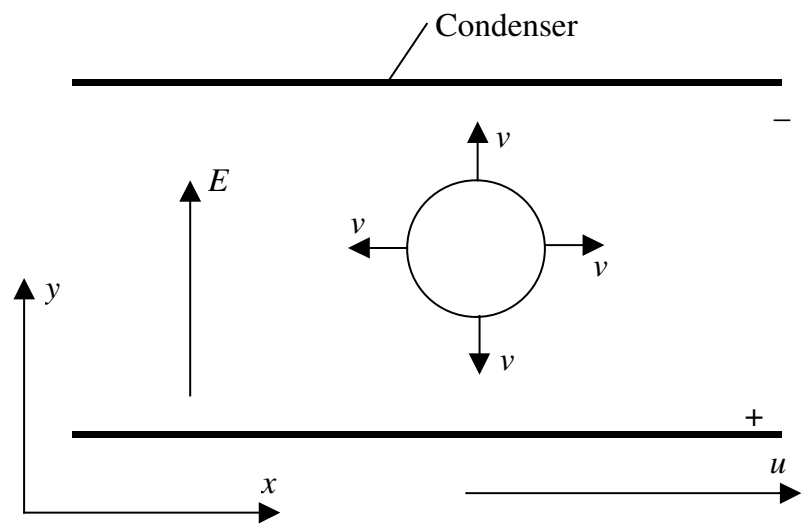

Fig. 1. Expanded circumference inside the parallel plate charged condenser